\pdfoutput=1 

\documentclass[10pt]{article}
\usepackage[margin=3.0cm]{geometry}
\usepackage{amsfonts}
\usepackage{amsmath}
\usepackage{amssymb}
\usepackage{amsthm}
\usepackage[sort&compress]{natbib}
\bibpunct{[}{]}{,}{a}{}{;}	% use this natbib punctuation
\usepackage{graphicx}
\usepackage{subfigure}
\usepackage{color}
\usepackage{url}
\usepackage{verbatim}		% for block LaTeX {comment} environment
\usepackage[bookmarks, colorlinks=false]{hyperref}
\usepackage{ulem}			% better underlining

\newcommand\LL{Little's law}
\newcommand\mrm{$M/M/1//p$}						% machine repairman (MRM)
\newcommand\dmm{$D/M/1//p$}						% post-service waiting room
\newcommand\gmm{$G/M/1//p$}						% robustness thm
\newcommand\mgm{$M/G/1//p$}						% state-dependent MRM

\def\figs{Figures}

\newtheorem{thm}{Theorem}
\newtheorem{con}{Conjecture}
\newtheorem{cor}{Corollary}
\newtheorem{lem}{Lemma}

\theoremstyle{remark}
\newtheorem{rem}{Remark}
\theoremstyle{definition}
\newtheorem{defn}{Definition}

\begin{document}

\title{A General Theory of Computational Scalability Based on Rational Functions}
\author{Neil J. Gunther\thanks{Performance Dynamics Company, 4061 East Castro Valley Blvd.,
Suite 110, Castro Valley, CA 94552, USA. Email:~{\tt nj gunther @ perfdynamics . com}}}
\date{\today}

\maketitle

\abstract
The {\em universal scalability law} of computational capacity is a
rational function \mbox{$C_p = P(p)/Q(p)$} with $P(p)$ a linear
polynomial and $Q(p)$ a second-degree polynomial in the number of
physical processors $p$, that has been long used for statistical
modeling and prediction of computer system performance. We prove that
$C_p$ is equivalent to the synchronous throughput bound for a
machine-repairman with state-dependent service rate. Simpler rational
functions, such as Amdahl's law and Gustafson speedup, are corollaries
of this queue-theoretic bound. $C_p$ is further shown to be both
necessary and sufficient for modeling all practical characteristics of
computational scalability.

%%%%%%%%%%%%%%%%%%%%%%%%%%%%%%%%%%%
\section{Introduction} \label{sec:intro}
For several decades, a class of real functions called {\em rational
functions}~\citep{ratfuns}, has been used to represent throughput
scalability as a function of physical processor configuration. In
particular, Amdahl's law~\citep{amdahl}, its modification due to
Gustafson~\citep{gusto} and the Universal Scalability Law
(USL)~\citep{cmg93} have found ubiquitous application. In this context, 
the relative computing capacity, 
$C_p$, is a rational function of the number of physical processors $p$. It
is defined as the quotient of a polynomial $P(p)$ in the numerator and
$Q(p)$ in the denominator, i.e., $C_p=P(p)/Q(p)$. Each of the
above-mentioned scalability models is distinguished by the number of
coefficients or fitting parameters associated with the polynomials in
$P(p)$ and $Q(p)$. For example, Amdahl's law and Gustafson's modification
are single parameter models, whereas the USL model contains two parameters.

Despite their historical utility, these models have stood in isolation
without any deeper physical interpretation. It has even been suggested
that Amdahl's law is not fundamental~\citep{prep}. More importantly, the
lack of a unified physical interpretation has led to the use of certain
flawed scalability models~\citep{arxiv02}. In this note, we demonstrate
that the aforementioned class of rational functions corresponds to
certain performance bounds belonging to a queue-theoretic model.

The idea that Amdahl's law, which has most frequently been associated with
the scalability of massively parallel systems, can be considered from a
queue-theoretic standpoint, is not entirely new~\citep[See
e.g.,][]{kleinrock92,nelson}. However, quite apart from motivations
entirely different from our own, those previous works employed {\em open}
queueing models with an unbounded number of requests (See
Appendix~\ref{sec:nelson}), whereas we shall use a {\em closed} queueing
model with a finite number of requests $p$ corresponding to the number of
physical processors. The USL function is associated with a
state-dependent generalization of the machine repairman~\citep{ipl}.

The organization of this paper is as follows. We briefly review the
scalability models of interest in Sect.~\ref{sec:fmodels}. The appropriate
queueing metrics associated with the standard machine repairman and its
state-dependent extension are discussed in Sect.~\ref{sec:qmodels}. The
performance characteristics associated with {\em synchronous} queueing are
also presented there. The main theorem (Theorem~\ref{eqn:mainthm}) is
established in Sect.~\ref{sec:qbounds}. Amdahl's law and Gustafson's linear
speedup are shown to be corollaries of this theorem. Finally, in
Sect.~\ref{sec:extrema} we prove an earlier conjecture that a rational
function with $Q(p)$ a second-degree polynomial is both necessary and
sufficient to model all practical cases of computational scalability.

%%%%%%%%%%%%%%%%%%%%%%%%%%%%%%%%%%%
\section{Parametric Models} \label{sec:fmodels}
Although technically, we are discussing rational functions, we shall
hereafter refer to them as parametric models, and the coefficients as
parameters, since the primary application of these models is nonlinear
statistical regression of performance 
data~\citep[See e.g.,][and references therein]{cmg93,gcap,pdcs,zone}.

\begin{figure}[!htb]
\centering
\includegraphics[scale = 0.75]{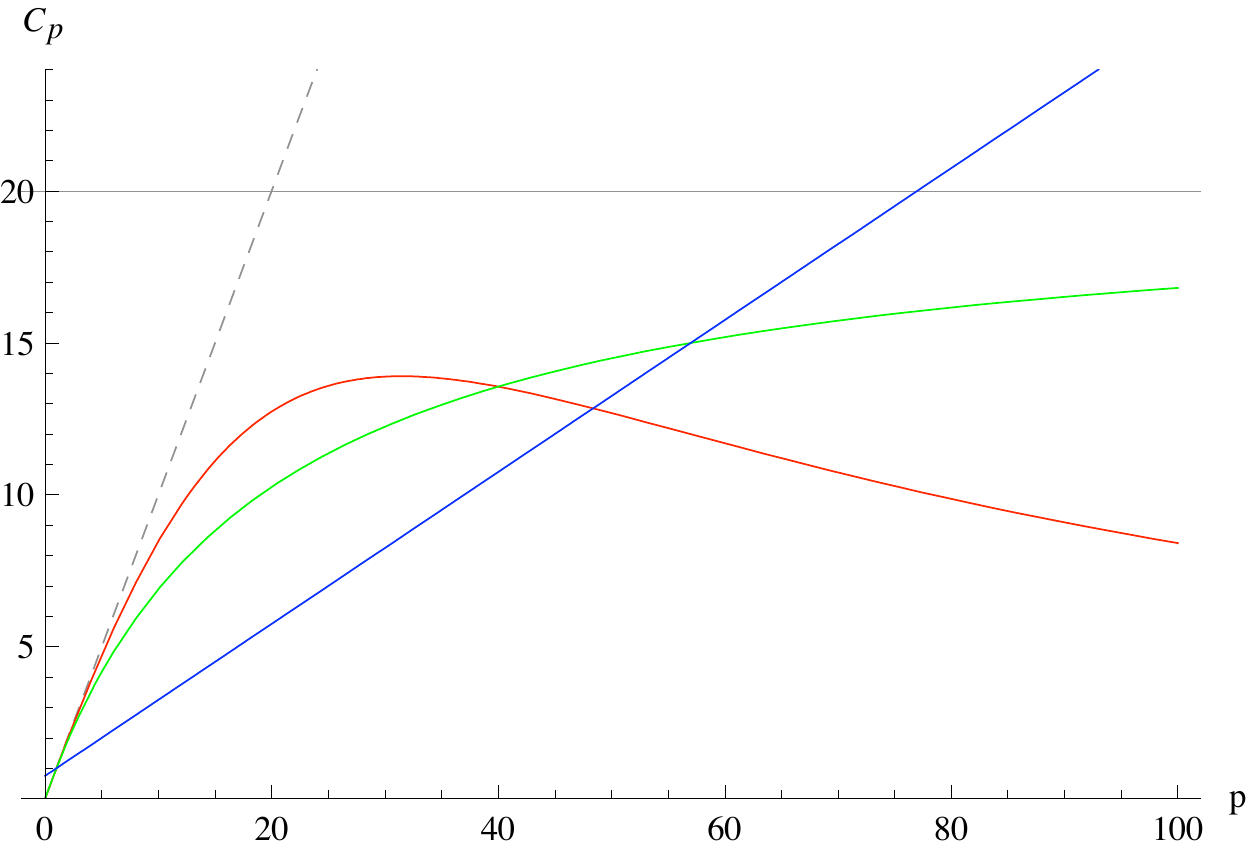} 
\caption{Parametric models: USL (red), Amdahl (green), Gustafson (blue),
with  parameter values exaggerated to distinguish their typical
characteristic relative to ideal linear scaling (dashed). The horizontal
line is the Amdahl asymptote at $\sigma^{-1}$}.
\label{fig:funcmodels}
\end{figure}  

\begin{defn}[Speedup] \label{def:speedup}
If an amount of work $N$ is completed in time $T_1$ on a
uniprocessor, the same amount of work can be completed in time $T_p < T_1$ on
a $p$-way multiprocessor. The speedup $S_p = T_1/T_p$ is one measure of 
scalability.
\end{defn}

\subsection{Amdahl's law} \label{sec:amdahl}
For a single task that takes time $T_1$ to execute on a uniprocessor ($p=1$),
Amdahl's law~\citep{amdahl} states that if the task can be {\em equipartitioned} onto 
$p$ processors, but contains an irreducible fraction of sequential work
$\sigma \in [0,1] \label{eqn:sigma}$,
then only the remaining portion of the execution time $(1-\sigma) T_1$ can 
be executed as $p$ parallel subtasks on $p$ physical processors.
The bound on the achievable equipartitioned speedup~\citep{ware} is given by the ratio
\begin{equation}
S_p(\sigma) = \dfrac{T_1}{\sigma T_1 + \biggl( \dfrac{1-\sigma}{p} \biggr) T_1}  \label{eqn:amdratio}
\end{equation}
which simplifies to
\begin{equation}
S_p(\sigma) = \dfrac{p}{1 + \sigma (p-1)} \, ; \label{eqn:amdahl}
\end{equation}
a rational function with $P(p)=p$ and $Q(p)$ a first-degree polynomial.
As the processor configuration is increased, i.e., 
$p \rightarrow \infty$, the number of concomitantly
smaller subtasks also increases and the speedup (\ref{eqn:amdratio})
approaches an asymptote,
\begin{equation}
S_p(\sigma)  \sim \sigma^{-1} \, ,  \label{eqn:amdasymp}
\end{equation}
shown as the horizontal in Fig.~\ref{fig:funcmodels}.

\subsection{Gustafson's speedup} \label{sec:gusto}
Amdahl's law assumes the size of the work is fixed. Gustafson's
modification is based on the idea of scaling up the size of
the work to match $p$. This assumption results in the
theoretical recovery of linear speedup
\begin{equation}
S^G_p(\sigma) = \sigma + (1-\sigma) p  \label{eqn:gusto}
\end{equation}
Equation (\ref{eqn:gusto}) is a rational function with $Q(p)=1$ and $P(p)$
a first-degree polynomial in $p$.

Although (\ref{eqn:gusto}) has inspired various efforts for improving
parallel processing efficiencies, achieving truly linear speedup has turned out to be
difficult in practice. Most recently, (\ref{eqn:gusto}) has been proposed
as a way to optimize the throughput of multicore processors~\citep{sutter}.

\begin{defn}[Relative Capacity] \label{def:scalability}
As an alternative to the speedup, 
scalability can also be defined as the relative capacity, $C_p = X(p)/X(1)$, 
where $X(p)$ is the throughput with $p$ processors, 
and $X(1)$ the throughput of a single processor.
\end{defn}

\subsection{Universal Scalability Law (USL)} \label{sec:usl}
The USL model~\citep{cmg93,pdcs,gcap} is a rational function 
with $P(p)=p$ and $Q(p)$ a second-degree polynomial:
\begin{equation}
C_p(\sigma,\kappa) = \dfrac{p}{1 + \sigma (p-1) + \kappa \, p (p-1)} \label{eqn:usl}
\end{equation}
where the coefficients belonging to the terms in the denominator 
have been regrouped into three terms with two parameters $(\sigma,\kappa)$.
These terms can be interpreted as representing:
\begin{enumerate}
\item Ideal concurrency associated with linear scalability (\mbox{$\sigma, \kappa = 0$})
\item Contention-limited scalability due to serialization or queueing (\mbox{$\sigma > 0, \kappa = 0$})
\item Coherency-limited scalability due to inconsistent copies of data (\mbox{$\sigma, \kappa > 0$})
\end{enumerate}
Table~\ref{tab:appclass} summarized how these parameter values can be used to  
classify the scalability of different types of applications.

Equation (\ref{eqn:usl}) subsumes (\ref{eqn:amdahl}) and
(\ref{eqn:gusto}). In particular, (\ref{eqn:amdahl}) is identical to 
(\ref{eqn:usl}) with $\kappa = 0$.
The key distinction is that, unlike (\ref{eqn:amdahl}), (\ref{eqn:usl}) 
possesses a maximum at
\begin{equation}
p^* = \sqrt{\dfrac{1-\sigma}{\kappa}}
\end{equation}
which is controlled by the parameter values according to:
\begin{enumerate}
\renewcommand{\labelenumi}{(\alph{enumi})}
\item $p^* \rightarrow 0$ as $\kappa \rightarrow \infty$ \label{item:kinf} 
\item $p^* \rightarrow \infty$ 			as $\kappa \rightarrow 0$ \label{item:kamd} 
\item $p^* \rightarrow \kappa^{-1/2}$ 	as $\sigma \rightarrow 0$ \label{item:khalf} 
\item $p^* \rightarrow 0$ 				as $\sigma \rightarrow 1$ \label{item:sigone}
\end{enumerate}

\noindent
The important implication is that beyond $p^*$ the throughput becomes {\em
retrograde}. See Fig.~\ref{fig:funcmodels}. This effect is commonly observed in applications that
involve shared-writable data (Case D in Table~\ref{tab:appclass}).

In the subsequent sections, we attempt to provide deeper insight into the
physical significance of (\ref{eqn:usl}) by recognizing its association
with queueing theory.

\begin{table} 
\caption{Application domains for the USL model} \label{tab:appclass}
\centering
\begin{tabular}{l|l} 
\hline
\multicolumn{1}{l|}{{\bf A: Ideal concurrency} ($\sigma,\kappa = 0$)} & 
 \multicolumn{1}{l}{{\bf B: Contention-limited} ($\sigma > 0, \kappa = 0$)} \\
\hline
Single-threaded tasks		& Tasks requiring locking or sequencing  \\
Parallel text search		& Message-passing protocols  \\
Read-only queries			& Polling protocols (e.g., hypervisors)  \\
\hline
\multicolumn{1}{l|}{{\bf C: Coherency-limited} ($\sigma = 0, \kappa > 0$)}	& 
 \multicolumn{1}{l}{{\bf D: Worst case} ($\sigma, \kappa > 0$)} \\
\hline
SMP cache pinging 		& Tasks acting on shared-writable data \\
Incoherent application state between  	& Online reservation systems \\
cluster nodes	& Updating database records  \\
\hline
\end{tabular}
\end{table}

%%%%%%%%%%%%%%%%%%%%%%%%%%%%%%%%%%%
\section{Queueing Models} \label{sec:qmodels} 
\begin{figure}[!htb]
\centering
\includegraphics[scale = 0.75]{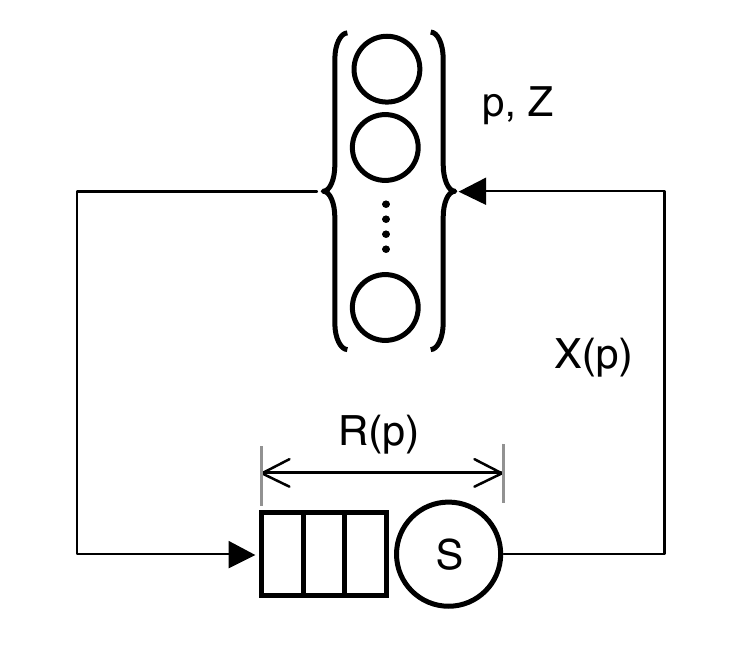} 
\caption{Conventional machine repairman queueing model comprising $p$ 
machines with mean uptime $Z$ ({\em top}) and a repair queue 
({\em bottom}) with mean service time $S$.} \label{fig:mrm}
\end{figure}

The {\em  machine repairman} (Fig.~\ref{fig:mrm}) is a well-known queueing model~\citep{gross} 
which represents an assembly line comprising a finite number of machines 
$p$ which break down after a mean lifetime $Z$.  A
repairman takes a mean time $S$ to repair a broken machine and if
multiple machines fail, the additional machines must queue for
service in FIFO order. The queue-theoretic notation, \mrm,\	implies 
exponentially distributed lifetimes and service periods with a finite
population $p$ of requests and buffering. 

\subsection{Queueing Metrics} \label{sec:qmetrics}
The performance characteristics of interest for the subsequent discussion 
are the throughput $X(p)$ and residence time $R(p)$.
\begin{defn}[Throughput]
The throughput, $X = N/T$, is the number of tasks $N$ 
completed in time $T$.
\end{defn}

\begin{defn}[Residence Time] \label{def:reztime}
The mean residence time $R = W + S$ is the sum of the time spent 
waiting for service $W$ plus the actual repair time $S$ 
when the repairman services the machine.
\end{defn}

On average, the 
number of machines that are ``up'' is $Z X$, while $Q$ are ``down'' (for repairs),
such that the total number of machines in either state is given by 
\mbox{$p = Q + ZX$}. Rearranging this expression produces:
\begin{equation*}
Q = p - ZX
\end{equation*}
and applying \LL~\citep{gross} 	($Q=XR$)~\citep{gross}
\begin{equation*}
XR = p - ZX
\end{equation*}
gives
\begin{equation}
R(p) = \dfrac{p}{X(p)} - Z   \label{eqn:mrmrt}
\end{equation}
for the mean residence time at the repair station.
Rearranging (\ref{eqn:mrmrt}) provides an expression for the mean throughput of the repairman 
as a function of $p$:
\begin{equation}
X(p) = \dfrac{p}{R(p) + Z} \label{eqn:mrmtput}
\end{equation}

\begin{defn}[Mean RTT] \label{def:meanRTT}
The denominator in (\ref{eqn:mrmtput}), $R(p)+Z$, is the mean {\em round-trip time} (RTT)  
for \mrm. 
\end{defn}

\begin{table}
\centering
\caption{Interpretation of the queueing metrics in Fig.~\ref{fig:mrm}} \label{tab:mrm}
\begin{tabular}{clll}	
		& \multicolumn{3}{c}{\bf Interpretation} \\
\cline{2-4}
\bf Metric & \bf Repairman & \bf Multiprocessor & \bf Time share\\
\hline
p 		& machines 		& processors			& users\\
Z 		& up time 			& execution period		& think time\\
S 		& service time		& transmission time 	& CPU time\\
R(p) 	& residence time 	& interconnect latency	& run-queue time\\
X(p) 	& failure rate    	& bandwidth				& throughput\\
\hline\end{tabular}
\end{table}

Since \mrm\	 is an abstraction, it can be applied to different
computational contexts. In the subsequent sections, the $p$ machines will
be taken to represent physical processors and the time spent at the repair
station is taken to represent the interconnect latency between the
processors~\citep{reed,balbo}. See Table~\ref{tab:mrm}. This choice is
merely to conform to the conventions most commonly used in discussions of
parallel scalability~\citep{pdcs}, but the generic nature of queueing model
means that any conclusions also hold for software 
scalability~\citep[See e.g.,][Chap. 6]{gcap}.

\subsection{Synchronous Queueing}
We consider the special case of {\em synchronous queueing} in \mrm. The 
queue-theoretic performance metrics defined in Sect.~\ref{sec:qmetrics} are 
for the steady-state case and therefore each corresponds to the statistical 
mean of the respective random variable. Moreover, as already mentioned, we 
cannot give an explicit expression for $R(p)$ since its value is dependent 
on the value of $X(p)$, which is also unknown in steady state.

\begin{defn}[Synchronized Requests] \label{def:syncing}
If all the machines in \mrm\	break down simultaneously, the queue
length at the repairman is maximized such that the residence time in
definition~\ref{def:reztime} becomes $R(p)=pS$. This situation corresponds
to one machine in service and $(p-1)$ waiting for service and provides 
a lower bound on (\ref{eqn:mrmtput})~\citep{muntz1,bb,muntz2}:
\begin{equation}
\dfrac{p}{pS + Z} \leq X(p)
\end{equation}
In the context of multiprocessor scalability (Table~\ref{tab:mrm}), it is
tantamount to all $p$ processors simultaneously issuing requests across the
interconnect.
\end{defn}

\begin{rem}[A Paradox] \label{rem:paradox}
Consider the case where all $p$ processors have the same deterministic $Z$ period.
At the end of the first $Z$ period, all $p$ requests will enqueue at the
interconnect (lower portion of Fig.~\ref{fig:mrm}) simultaneously. By
definition, however, the requests are serviced serially, so they will
return to the parallel execution phase (top portion of Fig.~\ref{fig:mrm})
separately and thereafter will always return to the interconnect at
different times. In other words, even if the queueing system starts with
synchronized visits to the interconnect, that synchronization is
immediately lost after the first tour because it is destroyed by the
serial queueing process. The resolution of this paradox is discussed in
Appendix~\ref{sec:syncing}.
\end{rem}

\begin{defn}[Synchronous RTT] \label{def:syncRTT}
In the presence of synchronous queueing, the mean RTT of 
definition~\ref{def:meanRTT} becomes $pS + Z$.
\end{defn}

\begin{lem} \label{lem:speedcap}
The relative capacity $C_p$ and the speedup $S_p$ give identical values for the
same processor configuration $p$.
\end{lem}

\begin{proof}
Let the uniprocessor throughput is defined as $X_1 = N/T_1$ 
and the multiprocessor throughput is $X_p = N/T_p$. Hence
\begin{equation*}
C_p = \dfrac{X(p)}{X(1)} = \dfrac{N}{T_p} \dfrac{T_1}{N} \equiv S_p \label{eqn:speedcap}
\end{equation*}
follows from definition~\ref{def:speedup}.
\end{proof}

\begin{lem}[Serial Fraction] \label{lem:seriality}
The serial fraction (\ref{eqn:sigma}) can be
expressed in terms of \mrm\	 metrics 
by the identity~\citep{arxiv02,pdcs,gcap}:
\begin{equation}
\sigma =  \dfrac{S}{S + Z} \rightarrow  
\begin{cases} 
0	& \text{as $S \rightarrow 0, \; Z =$ const.},\\ 
1	& \text{as $Z \rightarrow 0, \; S =$ const}. 
\end{cases} \label{eqn:ident}
\end{equation}
If $\sigma = 0$, then there is no communication between processors 
and the interconnect latency therefore vanishes (maximal execution time).
Conversely, if $\sigma = 1$, then the execution time vanishes 
(maximal communication latency).
\end{lem}

\begin{proof}
The RTT for a single (unpartitioned) task in Fig.~\ref{fig:mrm} is $T_1=S+Z$.
The RTT for a $p$ equipartitioned subtasks is $T_p=p(S/p)+Z/p$.
From definition~\ref{def:speedup}, the corresponding speedup is 
\begin{equation}
S_p = \dfrac{S+Z}{S+Z/p} \label{eqn:mrmspeed}
\end{equation}
Equating (\ref{eqn:mrmspeed}) with (\ref{eqn:amdratio}), we find
\begin{equation}
S=\sigma T_1 \quad \text{and} \quad Z=(1-\sigma)T_1 \, . \label{eqn:SZ}
\end{equation}
Eliminating $T_1$ produces
\begin{equation}
S = \biggl( \dfrac{\sigma}{1-\sigma} \biggr) Z
\end{equation}
which, upon solving for $\sigma$, produces (\ref{eqn:ident}).
\end{proof}

See Appendix~\ref{sec:syncing} for another perspective.

\begin{defn} \label{def:servratio}
The quantity
\begin{equation}
\frac{Z}{S} = \frac{1-\sigma}{\sigma}
\end{equation}
is the {\em service ratio} for the \mrm\	 model.
\end{defn}

\begin{thm}[Speedup Duality] \label{lem:duality}
Let $(\sigma, \pi)$ be a continuous dual-parameter pair with $\sigma$ is
the serial fraction (\ref{eqn:ident}) and $\pi = 1-\sigma$. The Amdahl
speedup (\ref{eqn:amdahl}) is invariant under scalings of $(\sigma, \pi)$
by $p$.
\end{thm}

\begin{proof}
Using definition~\ref{def:servratio}, theorem~\ref{lem:duality} can be
represented diagrammatically as:
\begin{equation}
\begin{matrix}
				&			& \pi/\sigma = Z/S	& 	& \\
				& \swarrow 	&		&	\searrow 	& \\
Z \mapsto Z/p 	& 			&		&				& Z \mapsto Z\\
S \mapsto S 	& 			&		& 				& S \mapsto pS\\
				& \searrow 	&		& 	\swarrow	& \\
				&			& S_p(\sigma) 	&				& \\
\end{matrix} \label{eqn:diagram}
\end{equation}

\noindent
The path on the left hand side of (\ref{eqn:diagram}) corresponds to reducing
the single task execution time by $p$ (subtasks) while the interconnect
service time remains unchanged. This follows from
definition~\ref{def:syncing}, $R=pS$, but the service time for each subtask
is also reduced to $S/p$. Hence, $R=S$. Conversely, the right hand path of
(\ref{eqn:diagram}) corresponds to $p$ tasks, each with unchanged execution
time $Z$, but scaled service time $R=pS$. Both paths result in Amdahl's law
(\ref{eqn:amdahl}), which can be seen by first rewriting
(\ref{eqn:mrmspeed}) in terms of the service ratio $Z/S$ (definition~\ref{def:servratio}):
\begin{equation}
S_p =  \dfrac{1+ Z/S}{1+\dfrac{1}{p} \biggl( \dfrac{Z}{S} \biggr)} \, . \label{eqn:amdsr}
\end{equation}

\begin{enumerate}
\renewcommand{\labelenumi}{(\alph{enumi})}
\item Interpreting the denominator in (\ref{eqn:amdsr}) as belonging to the
left hand path of (\ref{eqn:diagram}) leads to the expansion
\begin{align*}
S_p &= \dfrac{S+Z}{\dfrac{S+Z}{p} + S - \dfrac{S}{p}}\\
	&= \dfrac{(S+Z)/S}{\dfrac{1}{p} \biggl( \dfrac{S+Z}{S} \biggr) + \dfrac{p-1}{p}}
\end{align*}
Collecting terms and simplifying produces:
\begin{equation}
S_p = \dfrac{p}{1 + \biggl( \dfrac{S}{S+Z} \biggr) (p-1)}
\end{equation}
which is identical to (\ref{eqn:amdahl}) upon substituting (\ref{eqn:ident}).

\item  Following the right hand path in (\ref{eqn:diagram}) leads to the
expansion
\begin{align*}
S_p &= \dfrac{p  S(1+Z/S)}{pS + Z}\\
	&= \dfrac{p}{ p \biggl( \dfrac{S}{S+Z} \biggr)  + \dfrac{Z}{S+Z} +\dfrac{S}{S+Z} -\dfrac{S}{S+Z}}
\end{align*}
Collecting terms in the denominator produces:
\begin{equation}
S_p = \dfrac{p}{ p \biggl(\dfrac{S}{S+Z}  \biggr)  + 1 -\dfrac{S}{S+Z}}
\end{equation}
which also yields (\ref{eqn:amdahl}) via (\ref{eqn:ident}).
\end{enumerate}
\end{proof}

\begin{rem}
Theorem~\ref{lem:duality} anticipates the interpretation of Gustafson's law
as a consequence of scaling the work size $Z \mapsto pZ$ in \mrm. See 
corollary~\ref{thm:gusto}.
\end{rem}

\subsection{State-Dependent Service} \label{sec:LDS}
We now consider a generalization of this machine repairman model in
which the residence time $R(p)$ includes an additional time that is
proportional to the load on the server, expressed as the number of
enqueued requests. Since the queue-length is a canonical measure of the
state of the system, the repairman becomes a {\em state-dependent}
server~\citep{ipl,gross}, denoted \mgm. Let the additional service time
be $S^{\prime}$ in the state-dependent progression:
\begin{align}
p	&=1:	&	R(1) 	&=1 S \nonumber \\
p	&=2:	&	R(2) 	&=2 \, (S + S^{\prime}) \nonumber \\
p	&=3:	&	R(3) 	&=3 \, (S + 2 \,S^{\prime})   \nonumber \\
p	&=4:	&	R(4) 	&=4 \, (S + 3 \,S^{\prime})   \label{eqn:LDR}  \\
	& 	\cdots	& 		&	\cdots \nonumber \\
p	&=p: 	&	R(p) 	&=p \, (S + (p-1) S^{\prime})	\nonumber  
\end{align} 
The extra time spent by each machine at the repair station increases linearly
with the additional number ``down'' machines. There is no stretching 
of the mean service time, $S$, when repairing a single machine.
\begin{rem}
In general, it is expected that $S^{\prime} < S$. It could, however, be a multiple 
of $S$, but that is clearly undesirable. 
\end{rem}

Some example applications of the state-dependent \mgm\	 model 
in a computational context include:
\begin{enumerate}
\renewcommand{\labelenumi}{(\alph{enumi})}
\item Pairwise Exchange: 
Modeling the performance degradation due to combinatoric pairwise exchange of data
between $p$ multiprocessor caches or cluster nodes. See
Sect.~\ref{sec:extrema}. 
\item Broadcast Protocol: 
If any processor broadcasts a request for data, the other $(p-1)$ processors
must stop and respond before computation can continue~\citep{pdcs}.
\item Virtual Memory:  
Each task is a program with its own working set of memory pages. Page
replacement relies on a higher latency device, such as a disk. As the
number of programs $p$ increases, page replacement latency causes the
system to ``thrash'' such that the throughput to become
retrograde~\citep{ipl}. Cf. Fig.~\ref{fig:mrm}.
\end{enumerate}

%%%%%%%%%%%%%%%%%%%%%%%%%%%%%%%%%%%
\section{Parametric Models as Queueing Bounds}  \label{sec:qbounds}
In this section, we show that the parametric scalability models in 
Section~\ref{sec:fmodels} correspond to certain throughput bounds on 
the queueing models in Section~\ref{sec:qmodels}.

\begin{thm}[Main Result] \label{eqn:mainthm}
The universal scalability law (\ref{eqn:usl})
is equivalent to synchronous relative throughput in \mgm.
\end{thm}

\begin{proof} \label{eqn:mainproof}
Let $S^{\prime}= c S$ in (\ref{eqn:LDR}), with $c$ a positive constant
of proportionality. The residence time for state-dependent, synchronous-requests becomes
\begin{equation}
R(p) = pS + c \, p(p-1)S    \label{eqn:LDrt}
\end{equation} 
Substituting (\ref{eqn:LDrt}) into definition~\ref{def:scalability}:
\begin{align}
C_p	 &= \dfrac{p (S + Z)}{pS + c  \, p(p-1) S + Z} \label{eqn:p2} \\
	 &= \dfrac{p (S + Z)}{(p-1)S + (S + Z) + c \, p(p-1)S } \nonumber\\
	 &= \dfrac{p (S + Z)}{(S + Z) \, [1 + (p-1)S(S + Z)^{-1} +  c \, p(p-1)S(S + Z)^{-1}]} \nonumber
\end{align}
Collecting terms and simplifying produces:
\begin{equation}
C_p = \dfrac{p}{1 + \sigma (p-1) +  c  \, \sigma p(p-1)} \, , \label{eqn:p1}
\end{equation}
where we have applied the identity for the serial fraction in lemma~\ref{lem:seriality}.
Combining the coefficients of the third term in the denominator of  
(\ref{eqn:p1}) as \mbox{$\kappa = c \,\sigma$}, yields (\ref{eqn:usl}). 
\end{proof}

\begin{rem}
Since $c$ is an arbitrary constant, $c > 0$ implies that the parameter 
$\kappa = c \,\sigma$ in (\ref{eqn:usl}) can be unbounded, whereas 
$\sigma < 1$ always.
\end{rem}

\begin{rem}
The state-dependence of $R(p)$ in (\ref{eqn:LDrt}) does not change
lemma~\ref{lem:seriality} since $\sigma$ is determined by $S$ and $Z$ only 
and both of those queueing metrics are constants.
\end{rem}

\begin{cor}[Amdahl's law] \label{thm:amdahl} 
Amdahl's law (\ref{eqn:amdahl}) is the synchronous bound on relative 
throughput in \mrm.
\end{cor}

\begin{proof}
Follows immediately from the proof of
theorem~\ref{eqn:mainthm} with $c=0$ in (\ref{eqn:p2}).
\end{proof}

\begin{rem}
Elsewhere~\citep{arxiv02,pdcs,gcap}, corollary~\ref{thm:amdahl}  was proven
as a separate theorem.
\end{rem}

\begin{cor}[Gustafson's law] \label{thm:gusto} 
Gustafson's law (\ref{eqn:gusto}) corresponds to the rescaling 
\mbox{$Z \mapsto pZ$} in \mrm.
\end{cor}

\begin{proof}
Since Gustafson's result is a modification of Amdahl's law, we start with
(\ref{eqn:p2}) and let $c=0$. Under $Z \mapsto p Z$, the scalability
function becomes:
\begin{align}
C_p	&= \dfrac{p (S + pZ)}{pS + pZ} \nonumber\\
	&= \dfrac{S + pZ}{S + Z}  \nonumber\\
	&= \dfrac{S + p(Z + S - S)}{S + Z}  \nonumber
\end{align}
Once again, after application of lemma~\ref{lem:seriality}, this simplifies to  
\begin{equation}
C_p	= \sigma + p - \sigma p \label{eqn:gusto1}
\end{equation}
which is identical to the linear speedup $S^G_p$ in (\ref{eqn:gusto}).
\end{proof}

\begin{rem}
Rewriting (\ref{eqn:gusto1}) as $C^{\prime}_p = p + \sigma/(1-\sigma)$, we
note that the additive constant $\sigma/(1-\sigma) = S/Z$ is the inverse of
the service ratio in definition~\ref{def:servratio}. In the context of
\mrm, rescaling the execution time, $Z \mapsto p Z$, prior to
partitioning, adds a {\em fixed} overhead ($pS/pZ$) to an otherwise linear
function of $p$, whereas the overhead in Amdahl's law is an {\em increasing}
function of $p$.
\end{rem}

The results of this section have also been confirmed using event-based simulations~\citep{zone}.

%%%%%%%%%%%%%%%%%%%%%%%%%%%%%%%%%%%
\section{Extrema and Universality} \label{sec:extrema}
Ideal linear scalability ($C_p \sim p, ~\text{for}~ p>0$) has a positive and
constant derivative. More realistically, large processor configurations ($p
\rightarrow \infty$) are expected to approach saturation, i.e., the
asymptote $C_{sat} \sim \sigma^{-1}$.

Any physical computational system that develops a scalability {\em maximum}
at $p_{*}$ in the positive quadrant ($p > 0$), means that $C_p$ must have
a negative derivative for $p > p_{*}$ and therefore relative
computational capacity falls below the saturation value or $C_p < C_{sat}$,
i.e., throughput performance becomes {\em retrograde}. Since this behavior
is undesirable, there is little virtue in characterizing the maximum beyond
the ability to quantify its location ($p_{*}$) using a given scalability
model. This observation leads to the following conjecture~\citep[See
also][p. 65]{gcap}, which we now prove.

\begin{con}[Universality] \label{thm:params}
For a rational function $C_p = P(p)/Q(p)$ with $P(p) = p$, a necessary and
sufficient condition for $C_p$ to be a model of computational scalability is, 
$Q(p) = 1 + a_1 \, p + a_2 \, p^2$ with coefficients $a_1, a_2 > 0$.
\end{con}
The simplest line of proof comes from considering latency rather 
than throughput. See Section~\ref{sec:remarks} for further discussion.

\begin{figure}[!htb]
\centering
\includegraphics[scale = 0.75]{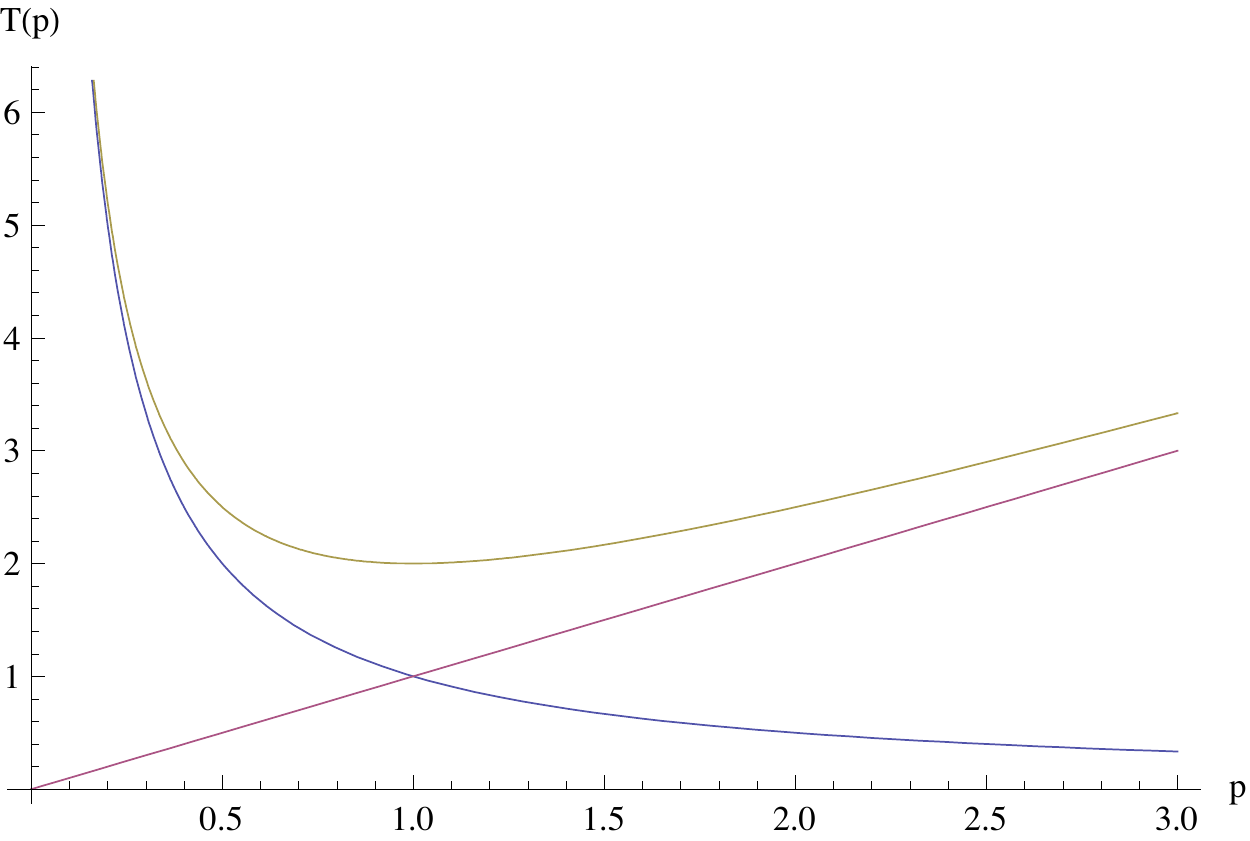} 
\caption{A minimum occurs in the total latency $T(p)$ due to an increasing
pairwise-exchange time being added to the initial latency reduction.}
\label{fig:latbrooks}
\end{figure}

\begin{proof}
Ideal latency reduction, $T_p = T_1/p$, is a hyperbolic function and
therefore has no extrema. The additional latency due to pairwise
interprocessor communication introduces the combinatoric term,
\mbox{$\binom{p}{2} = p(p-1)/2$}, such that the total latency becomes
\begin{equation}
T_p(\kappa) = \dfrac{T_1}{p} + \kappa \dfrac{T_1}{2}  (p-1) \label{eqn:brooks}
\end{equation}
with constant $\kappa > 0$. 
Equation(\ref{eqn:brooks}) has a unique minimum for \mbox{$p>0$}
(Fig.~\ref{fig:latbrooks}). Substituting (\ref{eqn:brooks}) into the
speedup definition~\ref{def:speedup}, produces
\begin{equation}
S_p(\kappa) = \dfrac{p}{1 + \kappa  p (p-1)}  \label{eqn:bscaling}
\end{equation}
where we have absorbed the factor of 2 in $\kappa$. 
$S_p(\kappa)$ now possesses a unique maximum in the positive quadrant ($p>0$). 
Thus, the quadratic term in the denominator of (\ref{eqn:bscaling}) is 
necessary for the existence of a 
maximum but it is not sufficient because $S_p(\kappa)$ does not exhibit  
the Amdahl asymptote $\sigma^{-1}$ when $\kappa = 0$.
However, the two-parameter latency 
\begin{equation}
T_p(\sigma, \kappa) = \dfrac{T_1}{p} + \sigma T_1 \biggl( \dfrac{p-1}{p} \biggr) + 
\kappa \dfrac{T_1}{2}(p-1) \label{eqn:quadtime}
\end{equation}
does introduce the required term into (\ref{eqn:bscaling}) and,  
by lemma~\ref{lem:speedcap}, is identical to (\ref{eqn:usl}). 
\end{proof}

The second term in (\ref{eqn:quadtime}) can be interpreted as the fixed
time it takes any one processor to broadcast a request for
data and wait for the remaining fraction of processors, \mbox{$(p-1)/p$}, to respond 
simultaneously. This is also another way to view synchronous queueing in \mrm.
It simply introduces a lower bound, $\sigma T_1$, on the latency reduction.

The third term in (\ref{eqn:quadtime}) is analogous to Brook's
law~\citep{brooks}: ``Adding more manpower to a late software project makes
it later,'' with $p$ interpreted as {\em people} rather than processors.
Here, for example, it can be interpreted as the latency due to the pairwise
exchange of data to maintain cache coherency in a multiprocessor.

\section{Conclusion}
Several ubiquitous scalability models, viz., Amdahl's law, Gustafson's law
and the universal scalability law (USL), belong to a class of rational
functions. Treated as parametric models, they are neither ad hoc nor
unphysical. Rather, they correspond to certain bounds on the relative
throughput of the machine repairman queueing model. In the most general
case, the main theorem~\ref{eqn:mainthm} states that the USL model
corresponds to the synchronous throughput bound of a load-dependent machine
repairman. USL subsumes both Amdahl's law and Gustafson's law
as corollaries of theorem~\ref{eqn:mainthm}. As well as providing a more physical
basis for these scalability models, the queue-theoretic interpretation has
practical significance in that it facilitates prediction of response
time scalability using (\ref{eqn:mrmrt}) and provides deeper insight
into potential performance tuning opportunities.

%%%%%%%%%%%%%%%%%%%%%%%%%%%%%%%%%
%%%%%%%%%%%%%%%%%%%%%%%%%%%%%%%%%

\appendix
\section*{Appendices}

%%%%%%%%%%%%%%%%%%%%%%%%%%%%%%%%%
\section{Why Universal?} \label{sec:universal}
The term  ``universal'' is intended to convey the notion that the USL
(defined in Sect.~\ref{sec:usl}) can be applied to {\em any} computer
architecture; from multi-core to multi-tier. This follows from the fact that
there is nothing in (\ref{eqn:usl}) that explicitly represents any
particular system architecture or interconnect topology. That information
is present but it is encoded in the numeric value of the parameters
$\sigma$ and $\kappa$.
The same could be said for Amdahl's law but the difference is that, 
being a rational function with linear $Q(p)$, Amdahl's law cannot
predict the retrograde scalability commonly observed in performance evaluation 
measurements~\citep{cmg93,gcap}. As proven in Sect.~\ref{sec:extrema}, the
USL is both necessary and sufficient to model all these effects.

The USL does not exclude defining a more general or more
complex scalability model to account for such details as, heterogeneous
processors or the functional form of degradation beyond $p^*$, but any such 
model must include the USL as a limiting case. The best
analogy might be to regard the USL as being akin to Newton's {\em universal
law of gravitation}. Here, ``universal'' means generally applicable to any
gravitating bodies. Newton's theory has been superseded by a more
sophisticated theory of gravitation; Einstein's general theory of
relativity. Einstein's theory, however, does not negate Newton's theory but 
rather, contains it as a limiting case, when space-time is flat. Since
space-time is flat in all practical applications, NASA uses Newton's
equations to calculate the flight paths of all its missions.

%%%%%%%%%%%%%%%%%%%%%%%%%%%%%%%%%
\section{Synchronous Queueing} \label{sec:syncing}
The proofs of theorem~\ref{eqn:mainthm} (USL) and
corollary~\ref{thm:amdahl} (Amdahl) employ mean value equations for 
metrics which characterize steady state conditions. As noted in
remark~\ref{rem:paradox}, synchronized queueing cannot be maintained in
steady state. Synchronization and steady state are not compatible concepts
because the former is an {\em instantaneous} effect,
whereas the analytic solutions we seek are only valid in long-run equilibrium.

Elsewhere~\citep{gcap,zone}, we have shown that a necessary requirement for
maintaining synchronous queueing is to introduce another buffer 
in addition to the waiting line ($W_1$) at the repairman. If
the extra buffer represents a \emph{post}-repair collection point, such that each repaired
machine (completed request) is held ``off-line'' until {\em all} $p$
machines are repaired then, synchronous queueing is maintained provided the
$Z$ periods are  i.i.d. deterministic. The extra buffer acts as a {\em barrier
synchronizer}. Unfortunately, this is a \dmm\	 queue, whereas the proof
of corollary~\ref{thm:amdahl} is based on \mrm. Moreover, the repairman
performance metrics are robust~\citep{bunday}, so our results should also
hold for \gmm.

In more recent simulation experiments~\citep{zone}, we have shown that this
restriction on $Z$ periods can be lifted by positioning the buffer as a
\emph{pre}-repair waiting room. Instead of requiring all $p$ machines to
break down and enqueue simultaneously, we allow any number, less than $p$,
to fail but with the added constraint that when {\em any} machine invokes
service at the repairman, all other machines (or executing processors) must
suspend operations as well, i.e., visit the suspension buffer. Under these conditions, $Z$
can be $G$-distributed. Because this {\em intermittent synchronization}
occurs with much higher frequency and for much shorter average time periods
than barrier synchronization, the potential impact of the $G$-distributed
tails on the $Z$ periods is truncated.

Synchronization can be treated as a two-state Markov process, e.g.,
$A$: {\em parallel} and $B$: {\em serial}, where the $B$ state 
includes those processes that are suspended as well as waiting for service.
If $\lambda_A$ is the transition rate for $A \to B$ and $\lambda_B$ for 
$B \to A$, then the probability of being in state $B$ is 
given by
\begin{equation}
\Pr(B) = \frac{\lambda_B}{\lambda_A + \lambda_B} \label{probB}
\end{equation}
In the previous scenario it only takes a single machine to fail to suspended 
all other machines. The failure rate is therefore 
$\lambda_A = 1/Z$ and the service rate is $\lambda_B = 1/S$. Substituting 
these into (\ref{probB}) produces and expression 
identical to the serial fraction $\sigma$ in (\ref{eqn:ident}). In
state $B$, some fraction $p_1$ are enqueued and the remainder $p_2 = p -
p_1$ are suspended. On average, any machine can expect to spend time $R =
(p_1 + p_2) S$ to complete repairs. Hence, the total serial time is
$R=pS$, which is the quantity that appears in the proofs.

%%%%%%%%%%%%%%%%%%%%%%%%%%%%%%%%%
\section{Queueing Models of Amdahl's Law} \label{sec:nelson}
Others have also considered Amdahl's law from a queue-theoretic
standpoint~\citep[See, e.g.,][]{kleinrock92,nelson}. Of these,
\citep{nelson} is closest to our discussion, so we briefly summarize the
differences.

First, the motivations are quite different. The author of~\citep{nelson},
like many other authors, seeks clever ways to \emph{defeat} Amdahl's
law; in the sense of Gustafson (Sect.~\ref{sec:gusto}), whereas we are
trying to \emph{understand} Amdahl's law by providing it with a more
fundamental physical interpretation. Ironically, both investigations invoke
queue-theoretic models to gain more insight into the pertinent issues; an
\emph{open} ($M/M/m$) queue in \citep{nelson}, a \emph{closed} queue 
(Fig.~\ref{fig:mrm}) in this paper.

Second, two steps are undertaken to define an alternative speedup function:
\begin{enumerate}
\item An attempt to unify both the Amdahl and Gustafson equations  
into a single speedup function.
\item Extend that unified speedup function to include waiting times.
\end{enumerate}
The overarching goal is to find waiting-time optima for this unified
speedup function. The unification step is achieved by purely algebraic
manipulations and does not rest on any queue-theoretic arguments. 
The open queueing model provides an ad hoc
means for incorporating waiting times as a function of queue length. The
subsequent analysis is based entirely on simulation results and thereafter 
departs significantly from the analytic approach of this paper.

By virtue of our approach, we have shown that both Amdahl and Gustafson 
scaling laws are unified by the {\em same} queueing model, viz., the machine-repairman
model. Moreover, corollary~\ref{thm:amdahl} is a lower bound on throughput; 
synchronous throughput, and therefore represents worst-case
scalability. With this physical interpretation, it follows
immediately that Amdahl's law can be ``defeated'' more conveniently than
proposed in~\citep{nelson} by simply requiring that all requests be issued
\emph{asynchronously}~\citep{zone}.

%%%%%%%%%%%%%%%%%%%%%%%%%%%%%%%%%
\section{Remarks on the Proof of Conjecture~\ref{thm:params}} \label{sec:remarks}
The proof in Section~\ref{sec:extrema}, is simplified by
using the {\em additive} properties of the latency function $T_p$
rather than inverting the rational function $f(p) = R(p)/Q(p)$ directly. 
Since $Q(p)$ is quadratic in $p$, the temptation is to consider the   
inverse of $f^{-1}$ and use the fact that a general quadratic function has a minimum. 

\begin{figure}[!htb]
\centering
\includegraphics[scale = 0.65]{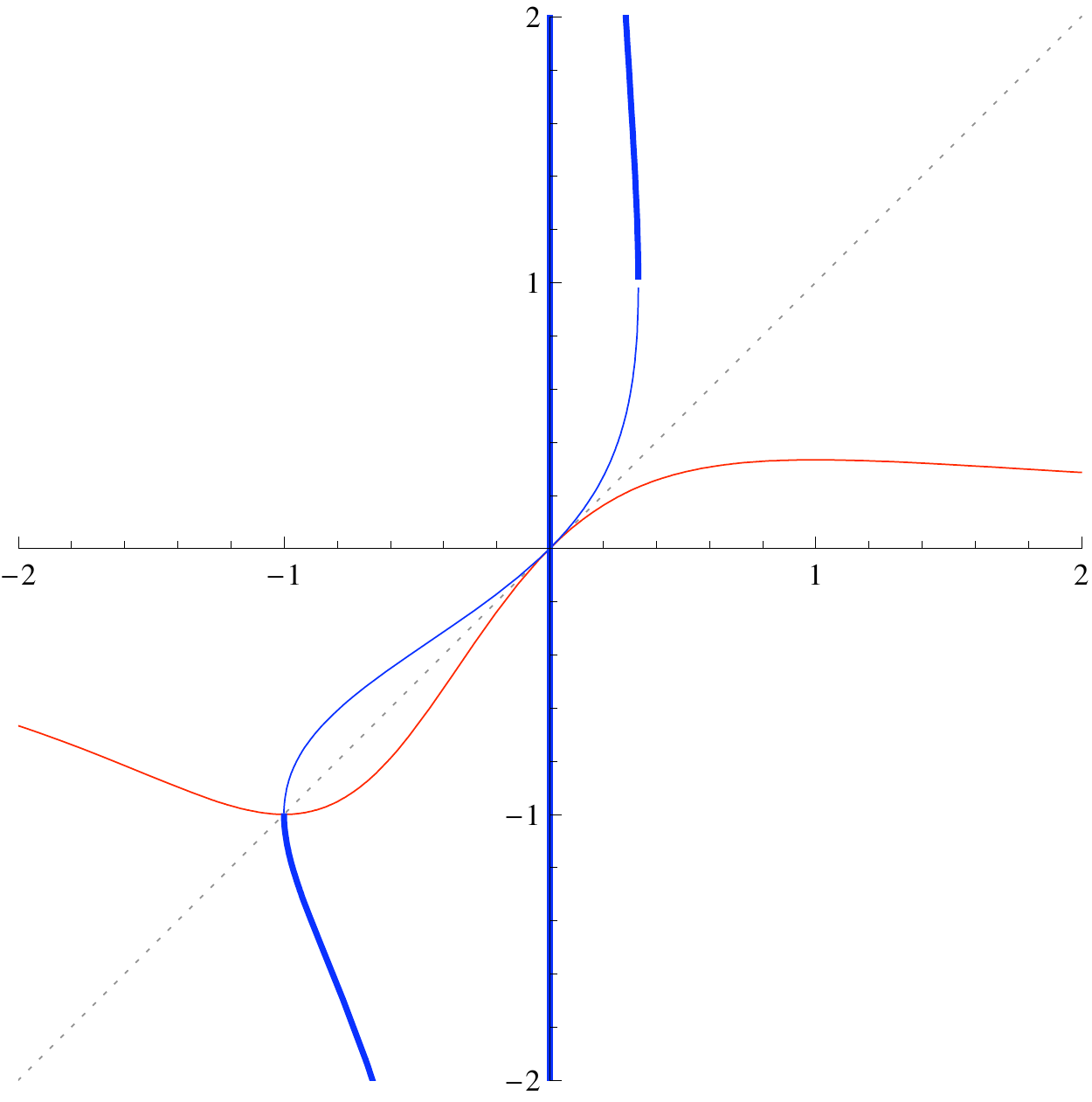} 
\caption{The complete rational function (\ref{eqn:simeq}) ({\em red}) and
its inverse ({\em blue}).}
\label{fig:uslratfun}
\end{figure}

To see why this approach runs into difficulties, 
consider the simplified representation of (\ref{eqn:usl}):
\begin{equation}
f(p) = \dfrac{p}{1 + p + p^2} \label{eqn:simeq}
\end{equation}
with all unit coefficients. Equation (\ref{eqn:simeq}) is not invertible in
the formal sense because $f^{-1}$ is not a one-to-one mapping, even in the
positive quadrant. We can, however, consider the full inverse with its
branches, as shown in Fig.~\ref{fig:uslratfun}.
The principal branch is shown in {\em light blue} with the other 
branches occurring at the extrema of $f^{-1}$. This corresponds to the 
extrema of (\ref{eqn:simeq}) occurring at $p = \pm 1 \Rightarrow f(1) = \pm 1/3$.
Hence, \mbox{$f^{-1} \in \mathbb{C}$}, \mbox{$\forall p > 1/3$}.

Alternatively, choosing the denominator be a perfect square:
\begin{equation}
\dfrac{p}{1 + 2p + p^2} \label{eqn:simsq}
\end{equation}
(\ref{eqn:simsq}) can be expressed either as a 
{\em product of linear factors}: 
\begin{equation*}
\dfrac{p}{1 + 2 p + p^2} = \biggl( \dfrac{p}{1+p}  \biggr) \biggl( \dfrac{1}{1+p}\biggr) 
\end{equation*}
or a {\em partial fraction expansion}:
\begin{equation*}
\dfrac{p}{1 + 2 p + p^2} =  \dfrac{1}{1+p} - \dfrac{1}{(1+p)^2}
\end{equation*}

\begin{figure}[!htb]
\centering
\includegraphics[scale = 0.75]{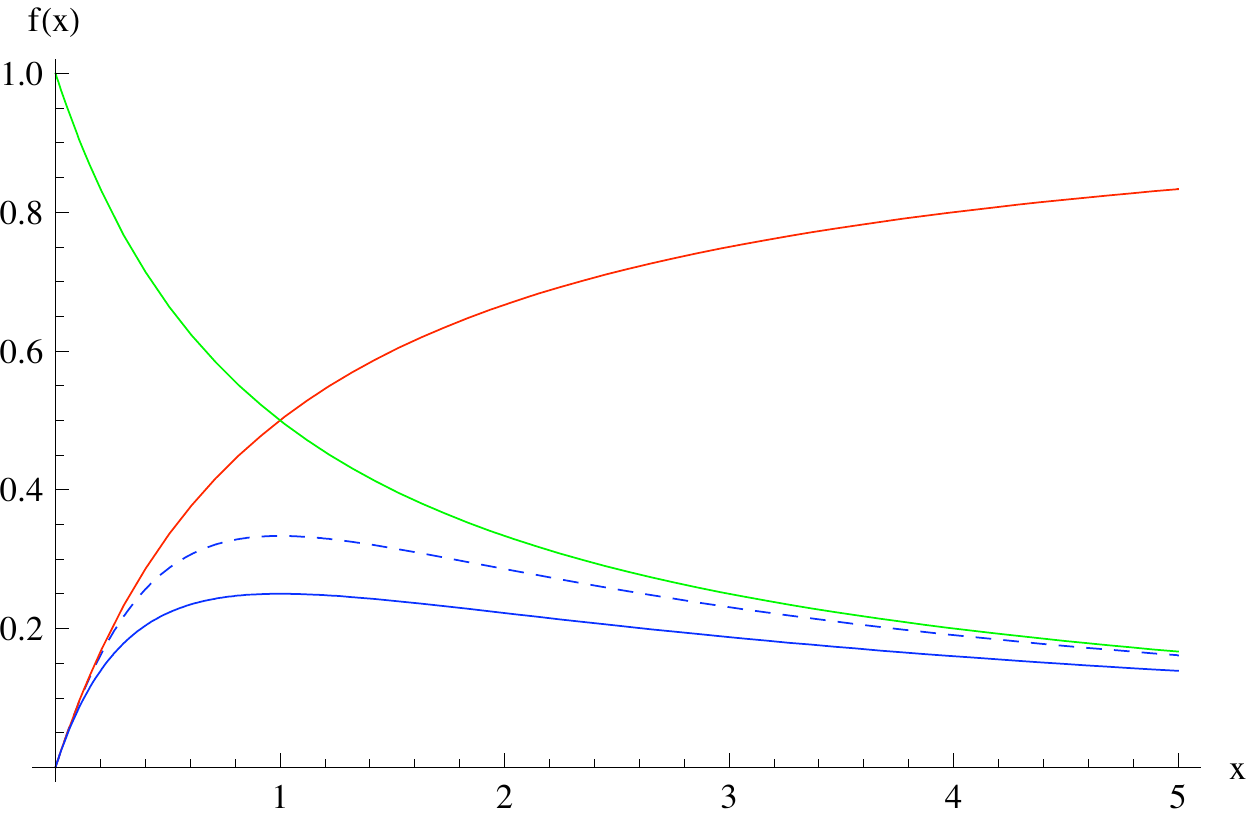} 
\caption{Amdahl scaling ({\em red}), envelope function $(1+p)^{-1}$ ({\em green}), 
their convolution ({\em solid blue}) and equation (\ref{eqn:simeq}) ({\em dashed blue}).
The small difference in the latter two curves arises from the factor of 2 in the 
denominator of equation (\ref{eqn:simsq}).}
\label{fig:uslconv}
\end{figure}  

Although such decompositions are suggestive of the need for two parameters
(Fig.~\ref{fig:uslconv}), they would seem to obscure the proof of
Theorem~\ref{thm:params} rather than illuminate it. 
Using the latency function $T_p$ and then
``inverting'' it to produce the corresponding throughput scaling using
lemma~\ref{lem:speedcap}, avoids these problems.

\section{Acknowledgments}
I thank Wen Chen for finding several typos in the original manuscript.

%%%%%%% REFERENCES %%%%%%%%%%
\setlength{\bibsep}{1pt}    % reduce vertical space b/w bibitems
\bibliography{uslthms}
\bibliographystyle{unsrt}

\end{document}